\documentstyle[12pt,epsf]{article} 

\renewcommand{\thefootnote}{\fnsymbol{footnote}}

\newcommand{\be}{\begin{equation}}
\newcommand{\ee}{\end{equation}}
\newcommand{\bea}{\begin{eqnarray}}
\newcommand{\eea}{\end{eqnarray}}

\begin{document}


\begin{titlepage}
\begin{flushright}
\begin{tabular}{l}
LPT - Orsay 99/32\\
ROMA  99/1248
\end{tabular}
\end{flushright}
\vskip3.5cm
\begin{center}
  {\Large \bf Comment on the heavy $\to$ light form factors\\}
  \vskip1cm {\large \sf Damir Becirevic$^{ \small {\mathrm a}}$, Alexei B. Kaidalov$^{ \small {\mathrm b}}$}
  \vskip0.5cm
$^{\mathrm a}$ Dip. di Fisica, Universit\`a ``La Sapienza" 
and INFN, Sezione di Roma,\\ 
P.le Aldo Moro 2, I-00185 Rome, Italy \\
  \vskip0.2cm
$^{\mathrm b}$ Institute of Theoretical and Experimental Physics,\\
 B. Cheremushkinskaya 25, 117526 Moscow, Russia\\
and Laboratoire de Physique Th\'eorique (B\^at. 210) \\
Universit\'e de Paris XI, 91405 Orsay Cedex, France\footnote{Unit\'e 
Mixte de Recherche C7644 du Centre National de 
la Recherche Scientifique.}\\
    \vskip1.8cm
  {\large\bf Abstract:\\[10pt]} \parbox[t]{\textwidth}{
  We propose a simple parametrization for the form factors in heavy to light 
  decays which satisfies the heavy quark scaling laws and avoids
  introducing the explicit ``dipole" form. The illustration is provided on the 
  set of lattice data for $B\to \pi$ semileptonic decay. The resulting shape of the form factor 
  also agrees with the QCD sum rule predictions.} 

\end{center}
\end{titlepage}

\renewcommand{\thefootnote}{\arabic{footnote}}
\setcounter{footnote}{0}

\noindent
{\bf 1.~\underline{Preliminaries}}
~~A lack of the precise information about the shapes of various form factors
from the first principles of QCD was, and still is the main source of 
uncertainties in extraction of the CKM parameters from the experimentally
measured exclusive decay modes~\cite{CLEO}. This problem is 
particularly pronounced in the {\sl heavy $\to$ light} processes because the 
kinematically accessible region is very large. 
In such a situation it is not only important to know the absolute value 
of a particular form  factor at one specific point (traditionally at $q^2=0$), 
but also its behaviour in $q^2$, which (when suitably integrated over the whole 
phase space) gives the physically 
measurable branching ratio. 
 So far, there were very many studies  
 devoted to that issue. In spite of the recent progress, there is
 still no method based on QCD only, which could be used to 
  describe the complicated nonperturbative dynamics in the whole physical 
  region\footnote{The recent reviews about the form factors 
  in heavy to light decays can be found in Refs.~\cite{LQCD,QCDSR}. They also contain an exhaustive list of references.}.  
Most of the approaches agree that the functional dependence of the 
{\sl heavy  $\to$ light}
 form factors is `distorted' in the region of not so high $q^2$ 
 (or equivalently, for
moderately and very large transfers, $\vec q^2$). In other words, it is 
different from 
what one would obtain by invoking the nearest pole dominance. 
Although somewhat intriguing, this discrepancy with the pole dominance hypothesis is however expected: the low $q^2$ region 
is very far from the first pole, so that 
many singularities (higher excitations and multiparticle states) 
become `visible' too, {\it i.e.} their contribution becomes sensible (comparable to that of the first pole). One would obviously like to quantify these effects of higher states.

For definiteness, we consider the form factors $f^{+,0}(q^{2})$ which parametrize
$B\to \pi \ell \nu$ decay as
\begin{eqnarray}
\label{gr}
\langle \pi (p')\vert V_\mu (0)\vert B(p)\rangle \ = \  
\biggl( p+p'-q \frac{m_B^2 -m_\pi^2}{q^2}\biggr)_{
\mu}\ f^+(q^2)\ +\ 
\frac{m_B^2 -m_\pi^2}{q^{2}} q_\mu \ f^0(q^2)\ ,
\label{def}
\end{eqnarray}
where $V_\mu=\bar u\gamma_\mu b$. $f^+(q^2)$ is the dominant form factor,
{\it i.e.} the only one needed for the decay rate 
\begin{eqnarray}
\frac{d\Gamma}{d q^2}(\bar B^{0} \rightarrow \pi^+ \ell^{-} \bar \nu_{
\ell}) \ =\ \frac{G_F^2 \vert V_{ub}\vert ^{2}}{192 \pi^3 m_B^3 }\
\lambda^{3/2}(q^2)\ \vert f^+ (q^2) \vert ^2 \ ,
\end{eqnarray}
in the case of a massless lepton\footnote{
$\lambda (t) = (t + m_{B}^{2} - m_{\pi}^{2})^{2} - 
4 m_{B}^{2} m_{\pi}^{2}$, is the usual triangle function.} ($\ell = e, \mu$). 
The subdominant form factor $f^0(q^{2})$, measures the divergence of the 
vector current and its contribution
is proportional to $m_\ell^2$ in the decay rate, thus it is irrelevant for
present experiments~\cite{CLEO}. Its knowledge is however important in lattice 
analyses, as well as in the approaches relying on the soft pion theorem 
(in which the form factors are calculated at large $q^2$), 
because 
 it can help to constrain the dominant $f^+(q^2)$, through  the kinematic 
 condition
 \bea
 f^+(0)\ = \ f^0(0)\quad .
\label{1st}
 \eea

\vspace{8mm}

\noindent
{\bf 2.~\underline{Constraints}}~~
Beside this important constraint~(\ref{1st}), in the heavy quark limit 
($m_B\to \infty$) and near the 
zero-recoil point ($\vec q ^2 \simeq 0$, or equivalently $q^2 \simeq 
q^2_{\rm max}$) where the pion is soft, one encounters the well known 
Isgur-Wise scaling law~\cite{IW}\footnote{
The overall logarithmic correction proportional to 
$\left( \alpha_s(m_b)/\alpha_s(\mu) \right)^{-2/\beta_0}$ (arising 
from the perturbative matching between the QCD and the 
heavy quark vector current) are irrelevant for our discussion.}
\bea
\langle \pi (p')\vert V_\mu \vert B(p)\rangle &\propto & 
\sqrt{m_B} \ \biggl( \ 1 + {\cal{O}}(1/m_B) \ \biggr) \; , \nonumber 
\eea
which in terms of form factors reads:
\bea
f^+(q^2\simeq q^2_{\rm max}, m_B)\sim \sqrt{m_B}\ ,~~~ \  
f^0(q^2\simeq q^2_{\rm max}, m_B)\sim {1\over \sqrt{m_B} }\ .
\label{22nd}
\eea
As first noticed in Ref.~\cite{Voloshin}, one may use the low energy theorem to get one more constraint, often invoked in $K_{\ell 3}$ decays. Namely, in the soft pion limit ($p'_\mu \to 0$ and
$m_\pi^2 \to 0$), the r.h.s. of~(\ref{def}) simply becomes $f^0(p^2) p_\mu$, 
while the l.h.s. from the LSZ reduction of the pion amounts to 
$\left( f_B/f_\pi \right) p_\mu$, so that one deduces:
\bea
f^0(m_B^2) = {f_B \over f_\pi}\; .
\label{2nd}
\eea
Next scaling law comes from the asymptotic limit of the light cone QCD sum rules 
which are applicable in the
low $q^2$ region. It was
first shown in Ref.~\cite{CHERNYAK}, that the form factors in heavy to light decays satisfy 
\bea
f^{+0}(q^2\simeq 0, m_B)\sim m_B^{-3/2} \quad  .
\label{counting}
\eea
This scaling law was recently derived also in the framework of the large energy effective theory~\cite{CHARLES}.
(\ref{counting}) will provide us an important additional constraint in what follows. 

Our knowledge of the form factors goes a little bit beyond. We know that the 
Lorentz invariant form factors are analytic functions satisfying
the dispersion relations:
\bea
f^{0}(q^2) &=& {1\over \pi} \int_{t_0}^\infty d t {{\mathrm Im} f^{0}(t)\over t - q^2 - i\epsilon} \quad , \\
&& \cr
f^+(q^2) &=&{ {\rm Res}_{q^2 = m_{B^*}^2} f^+(q^2)  \over  \textstyle{\large m_{B^*}^2} \ -\ 
\textstyle{\large q^2}}
 + {1\over \pi} \int_{t_0}^\infty d t {{\mathrm Im} f^{+}(t)\over t - q^2 - i\epsilon}\quad .
\label{DRel} 
\eea
The imaginary parts are consisting of all single and multiparticle 
states with the same quantum numbers as 
$f^+(t)\ (f^0(t))$, {\it i.e.} $J^P = 1^-\ (0^+)$, thus a multitude of poles and cuts situated above the $B\pi$ production threshold, $t_0=(m_B + m_\pi)^2$. 
However, below $t_0$ (but above the
semileptonic $B\to \pi$ region) there is a
$B^*$ vector meson ($m_{B^*}=5.325$~GeV) contributing to the $f^+(t)$ form factor. The residue at that pole is the product of the $B^* B\pi$ coupling and the
coupling of $B^*$ to the vector current:
\bea
{\rm Res}_{q^2 = m_{B^*}^2} f^+(q^2) = {1\over 2} m_{B^*} f_{B^*} g_{B^*B\pi}\ ,
\eea 
where the standard definitions were used:
\bea
\langle 0 \vert V_\mu \vert B^*(p,\epsilon ) \rangle &=& f_{B^*}m_{B^*}\epsilon_\mu\ ,\cr
&&    \cr
\langle B^-(p) \pi^+(q) \vert B^{*0}(p+q,\epsilon ) \rangle &=&  g_{B^*B\pi} (q \cdot
\epsilon)\ .
\eea 
While the lattice is providing us with better and better estimate
for $f_{B^*}$~\cite{HLAT}, the value of $g_{B^*B\pi}$ remains vague. 
From the above definitions, one immediately finds that ${\rm Res}_{q^2 = m_{B^*}^2}
f^+(q^2)$ scales as $m_B^{3/2}$.

The simple pole dominance would mean that only the first term in the r.h.s. of~(\ref{DRel}) survives, whereas all the other states which 
can couple to the vector current and are above $t_0$, would eventually cancel.
Moreover, one would encounter the discrepancy when trying
to reconcile the pole dominance ansatz with~(\ref{counting}), because the pole 
dominance gives $f^+(0)\sim m_B^{-1/2}$. 

The easiest way out was to suppose the pole behavior for $f^0(q^2)$, and the double pole (dipole) one
for $f^+(q^2)$~\cite{UKQCD,ALAIN}:
\bea
f^0(q^2)\ =\; \frac{f (0)\;}{\, 1 \ -\  q^2/m_{0^+}^2\, } \,, \quad \  \; f^+(q^2)\ =\; \frac{f (0)\;}{\left( \, 1 \ -\  q^2/m_{1^-}^2\, \right)^2 } \,. 
\eea
 This choice basically satisfies all constraints but it is 
not informative at all: we know that we have a pole at $B^*$ (and more importantly, 
we know its position) which we trade for a better fit with data,
without getting any supplementary physical information. For that reason, we
wanted to propose a new parametrization which would satisfy all the constraints
mentioned above, and use the fact that we know where the first pole is. In this way the fit with data will turn to be far more informative. 

\vspace{8mm}

\noindent
{\bf 3.~\underline{Proposal}}~~
One should always start from the fact that $f^+(q^2)$ {\bf  does
have} a pole at $m_{B^*}^2$. 
The contribution
of other higher states (for which we do not know positions, 
nor residues), 
can be parametrized as an effective pole whose position is
to be determined relatively to the first pole. In other words, we propose:
\bea
f^+(x) \ = \  c_B \ \left( 
{1\over 1 - x} - 
{\alpha \over 
1 -\frac{\textstyle{\large x}}{ \textstyle{\large \gamma}}
} 
 \right) \quad ,
\label{PROP}
\eea
where we introduced $x = q^2/m_{B^*}^2$, and parameters 
$\alpha$ and $\gamma$ - positive constants which scale with $m_B$. Obviously,
$c_B m_{B^*}^2$ is the residue of the form factor at $B^*$.
Note that $\gamma m_{B^*}^2$ is the squared mass of an effective $1^-$ excited $B^{*\prime}$-state, which is equal to $(m_{B^*} + \Delta)^2\simeq m_{B^*}^2  + 2 \Delta m_{B^*}$, where $\Delta$ does not depend on the heavy quark in the heavy quark limit. Therefore, $\gamma \simeq 1 + 2\Delta/m_{B^*}$.
We now check the scaling laws. At $q^2_{max}$, one has:
\bea
\label{scal2}
f^+ (x_{max})\simeq c_B m_B \left( 1 + {\alpha \gamma \over 
(1 - \gamma) m_B} \right)
\eea
Since $c_B$ scales as $\sim m_B^{-1/2}$, then in order to fulfill the heavy quark
 scaling law~(\ref{22nd}) $f^+ (x_{max}) \sim m_B^{1/2}$, $1 - \gamma$ can at most scale as $1/m_B$. That is consistent with the above observation on $\gamma$.  
It is also clear that in this counting, $\alpha \gamma$ scales as a constant. 
On the other hand, at $q^2=0$ we have:
\bea
f^+(0) = c_B (1 - \alpha) \sim m_B^{-1/2} (1 - \alpha) \sim m_B^{-3/2}\quad ,
\eea
which means that $(1 - \alpha)$ scales as $\sim 1/m_B$. This shows why we 
have chosen the sign ``$-$'' between the two terms in (\ref{PROP}). 
In fact, it is easy to see that both $\alpha$ and $\gamma$ scale as 
$1 + {\rm {\small const.}}/m_B + \cdots$.  To build $f^0$, we take 
into account $f^0(0)=f^+(0)$. For convenience, we can measure 
the position of the effective pole to $f^0$ in terms of $x=q^2/m_{B^*}^2$, and write
\bea
f^0(x)={c_B (1 - \alpha ) \over \quad 1\ -\ \frac{\textstyle{\large x}}{ \textstyle{\large \beta}}\quad }\quad .
\label{f0}
\eea
One can easily see that $f^0(q^2_{max})\sim m_{B}^{-1/2}$ as it should, 
while $f^0(0)\sim m_{B}^{-3/2}$ by construction. Using the same argument as for $\gamma$, it is obvious that $\beta$ also scales like $1 + {\cal O}(1/m_B)$.

These four parameters ($c_B$, $\alpha$, $\beta$, $\gamma$) contain a considerable
information on the process: $c_B$ gives a residue of the form factor at the pole $B^*$,
$\alpha$ measures the contribution of the higher states which are parameterized by
an effective pole at $m_{B^*_{\mathrm eff}}^2 = \gamma \cdot m_{B^*}^2$. Contributions to the scalar form
factor are parameterized by another effective pole whose position is at 
$m_{B^*_{0\ \mathrm eff}}^2 =\beta \cdot m_{B^*}^2$. The situation with the
 form factor $f^+$ resembles the situation with the electromagnetic form factor
 of a nucleon, where the simplest $\rho$-dominance is not sufficient to describe
 a dipole behavior of the form factor at large $q^2$, and  one needs to
  introduce at least one extra excited vector meson.

 Present lattice analyses have limited statistical
accuracy and
 accessible kinematical region for the form factors and for a description of
 lattice data
a more constrained parametrization is required. In the limit of large energy of
 the light meson (large recoils), in the rest frame of the heavy meson 
\bea
E=\frac{m_B}{2}\left( 1-\frac{q^2}{m_B^2} + \frac{m_\pi^2}{m_B^2} \right) \  \stackrel{m_B\to \infty}{\simeq} \; \frac{m_B}{2}\left( 1 - x \right) \; .
\eea
In Ref.~\cite{CHARLES}, it was shown that in the heavy quark limit and for large recoils ($m_B\to\infty$ and $E\to\infty$), there exists one nontrivial relation between the two form factors, namely:
\bea
f^0=\frac{2E}{m_B}\,f^+\ .
\label{leet}
\eea
In this limit, and by using the above counting argument with $\alpha = 1 - \alpha_0/m_B$, $\gamma = 1 + \gamma_0/m_B$, and $\beta = 1 + \beta_0/m_B$, we have 
\bea 
&&f^+ = {c_0\ \alpha_0 \over \, 2 E \sqrt{m_B}\, }\  \left[ 1 - {\gamma_0 \over \alpha_0}\left( 1 - {m_B \over 2 E} \right) + {\cal O}(1/m_B) + \dots \right]\cr
&&\cr
&&f^0={c_0\ \alpha_0 \over \, 2 E \sqrt{m_B}\, } \ \left[ 1 + {\cal O}(1/m_B) + \dots \right]\; .
\eea 
Therefore, to satisfy the relation (\ref{leet}) it is necessary to 
 have $\gamma_0 = \alpha_0$. It is reasonable  then to take, $\alpha + \gamma = 2$, but also $\alpha = 1/ \gamma$, which satisfy (\ref{leet}) up to terms ${\cal O}(1/m_B^2)$. In what follows, we use $\alpha = 1/ \gamma$.

To summarize, the recipe is the following one: in lattice analyses, 
one first calculates $m_{B^*}$ (from the corresponding two-point correlation 
function)\footnote{In experiment, one fixes it to the physical mass of the vector meson $B^*$.}
to measure $x$, and then fits the data according to:
\begin{center}
\framebox[11cm][t]{\hspace*{-2.0cm}
\parbox[t]{15.3cm}{
\bea
{1\over f^0(x)} &=& {1 \over c_B (1 - \alpha )}  \ \left( 1 - \frac{\textstyle{\large x}}{ \textstyle{\large \beta}}\right) \; ,\cr
&& \cr
{1\over f^+(x)} &=& {1 \over c_B (1 - \alpha )}  \ \left( 1 - x \right)  \ 
\left( 1 - \alpha x \right) \; ,
\label{PARAMETR} 
\eea
}
\hspace*{2cm} }\\
\end{center}
\noindent
which makes altogether three parameters ($ c_B$, $\alpha$, $\beta$) to perform 
this fit. 

\vspace{8mm}

\noindent
{\bf 3.~\underline{Illustration}}~~
 Now, we would like to show how it works on the specific example with the set 
 of lattice data. Before the new (systematically improved) results are released\footnote{APE 
 and UKQCD groups, are already analyzing their new respective data.}, we decided
 to use the most recent 
 available results, those of JLQCD~\cite{JLQCD}.
\begin{table}[h!]
\begin{center}
\begin{tabular}{|c||c|c|c|c|c|c|}
\hline
\phantom{\Huge{l}} \raisebox{.1cm}{\phantom{\Huge{j}}}
$q^2$~[GeV$^2$]  & 20.7 & 19.2 & 18.3 & 17.4 & 16.4 & 15.4  \\  \hline
\phantom{\Huge{l}} \raisebox{.1cm}{\phantom{\Huge{j}}}
$x$  & 0.70 & 0.65 & 0.62 & 0.59 & 0.55 & 0.52  \\ \hline \hline
\phantom{\Huge{l}} \raisebox{.1cm}{\phantom{\Huge{j}}}
$f^{0}_{B\to \pi}(q^2)$ & 0.72(3)  & 0.71(3) & 0.67(2) & 0.66(4)  & 0.65(4) & 0.60(7) \\ \hline 
\phantom{\Huge{l}} \raisebox{.1cm}{\phantom{\Huge{j}}}
$f^{+}_{B\to \pi}(q^2)$ & 2.09(8)  & 1.52(8) & 1.41(7) & 1.28(10)  & 1.20(12) & 1.11(17)  \\ \hline 
\end{tabular}
\end{center}
\caption {\it Lattice results obtained by the Hiroshima group, JLQCD~{\rm \cite{JLQCD}}.}
\end{table}
\noindent
Since our purpose is only to illustrate the parametrization~(\ref{PARAMETR}),
 we will not further comment on the
technical details used in the calculation of JLQCD. However, one should
 note that
the lattice groups reported results by assuming the negligible 
SU(3) breaking effects, which is also supported by the QCD sum rule 
calculation~\footnote{Note that the LCQCD sum rules predict larger SU(3) breaking~\cite{PBall}, which is the consequence of the normalization of the light meson wave functions (implying that $f^+_{B\to K}(0)/f^+_{B\to \pi}(0)\simeq f_K/f_\pi$). This is in conflict with the findings of lattice calculations~\cite{LQCD}. Further research clarifying that issue is needed.}~\cite{Narison}, as well as by experiment with $D$-mesons~\cite{CLEO2}. 
That means that the pion used in the JLQCD calculation is consisted 
of degenerate quarks of mass near the strange one. Also, the spectator 
quark in $B$ meson is the $``s"$ quark\footnote{To our knowledge, among recent studies, only the APE 
collaboration~\cite{APE} made the extrapolation to the chiral limit.}.

\begin{figure}
\vspace*{-1.9cm}
\centerline{\epsfxsize12.0cm\epsffile{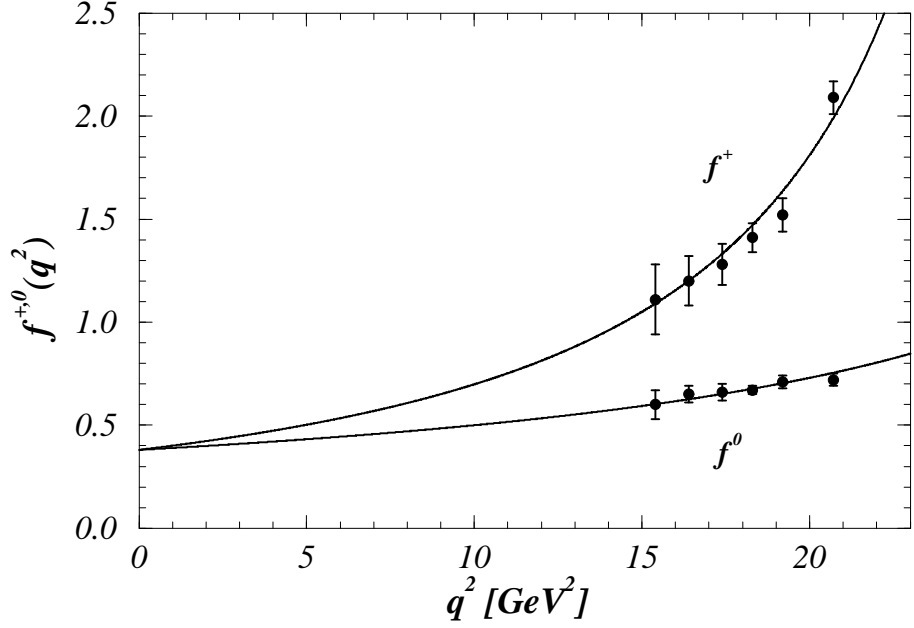}}
\vspace*{-0.6cm}
\caption[]{\it Fitting the lattice data using the parametrization~{\rm (\ref{PARAMETR})}.
Note that the fit of $f^+$ form factor is constrained by the precise
data for $f^0$. For easiness, only the central curves (without errors 
in parameters)
are displayed.}
\label{FIG1}
\end{figure}

First, we fix the mass of $m_{B^*} = 5.46$ GeV, as obtained in the same 
calculation of JLQCD. In realistic situation, one of course takes 
the experimental $m_{B^*} = 5.32$ GeV. Then we fit the data with the expressions~(\ref{PARAMETR}),
{\it i.e.} by minimizing:
\bea
\chi^2 &=& \sum_{i} \biggl\{ \biggl( {  
\left( {1/ f^0}\right)_i - {{ \textstyle{\large 1}}\over { \textstyle{\large c_B (1 - \alpha )}}}\left( 1 - \frac{\textstyle{\large x_i}}{ \textstyle{\large \beta}} \right)
 \over \delta \left( {1/ f^0} \right)_i } \biggr)^2  
 \ + \ 
\biggl( {  
\left( {1/[f^+ (1 - x)]}\right)_i - {{ \textstyle{\large 1}}\over { \textstyle{\large c_B (1 - \alpha  )}}}\left( 1 - \alpha x_i \right)
 \over \delta \left({1/[ f^+ (1 - x)]}\right)_i } \biggr)^2 \biggr\} \nonumber
\eea
from which we obtain the following result:
\bea
\label{res1}
f^{+,0}(0) = 0.38(8)\ ,\quad \alpha=0.54(17)\ ,\quad \beta=1.4(9)\ .
\label{RESULT}
\eea
The best fit and the lattice data are depicted in Fig.\ref{FIG1}.
These conservative errors in (\ref{RESULT}) are estimated from 
the correlated fit, 
corresponding to 
$\Delta \chi^2 = 3.53$ (68\% C.L.)~\cite{num_rec}, which defines 
the ellipsoid of errors (three parameters).
The errors we quote in (\ref{RESULT}) correspond to the extrema of that ellipsoid. 
As we already mentioned, 
by our parametrization, besides the first pole in $f^+(x)$ all the other singularities
are replaced by an effective pole, which
is situated at $m_{B^*_{\mathrm eff}}= m_{B^*} \sqrt{\gamma} = 
1.3^{+0.3}_{-0.2}\ m_{B^*}$. The effective pole for $f^0(x)$ is at 
$m_{B^*_{0\ \mathrm eff}}= m_{B^*} \sqrt{\beta} = 1.2^{+0.3}_{-0.5} \ m_{B^*}$.

An important advantage of this parametrization is its  transparent physical meaning. 
For example, after identifying $f(0)=c_B (1 - \alpha )$, we may also extract 
the information on 
the residue of the form factor $f^+$ at $B^*$, {\it i.e.} about the coupling
$g_{B^*B\pi}$:
 \bea
 c_B\ = 0.8(3) \quad \Rightarrow \quad g_{B^*B\pi} = 1.6(6) m_{B^*}/f_{B^*} \quad .
 \eea 
If we take the physical value $m_{B^*}=5.32$ GeV, and $f_{B^*}\simeq 0.2$ GeV,
we see that the resulting $g_{B^*B\pi} = 42\pm 16$, which is compatible with most of today's estimates of this coupling (see Tab.2 in Ref.~\cite{QCDSR}). A statistical (and more importantly, a systematic) improvement of the lattice data will provide more reliable estimate of this coupling.

The ellipsoid of errors is of the cigar shape so that one can reduce the errors and
consequently get some additional information, by
making reasonable assumptions. One can for instance make
 the following exercise. We saw that the position of the {\sl effective} pole 
 of the $f^+(q^2)$ function 
 is at $1.3^{+0.3}_{-0.2} \ m_{B^*}$. 
 If we now fix it to $(1.15, 1.20,1.25,1.30,1.35)\times m_{B^*}$,
 we obtain the results listed in Tab.~2.
\begin{table}[h!]
\label{FIT9}
\begin{center}
\begin{tabular}{|c||c c c c c|} \hline
\phantom{\Huge{l}} \raisebox{.1cm}{\phantom{\Huge{j}}}
$m_{B^*_{\mathrm eff}}/m_{B^*}$   & 1.15 & 1.20 & 1.25& 1.30 & 1.35     \\ 
\phantom{\Huge{l}} \raisebox{.1cm}{\phantom{\Huge{j}}}
$\alpha$  & 0.76 & 0.69 & 0.64 & 0.59 & 0.55   \\ \hline
\phantom{\Huge{l}} \raisebox{.1cm}{\phantom{\Huge{j}}}
$f(0)$   & 0.30(1)  & 0.32(1) & 0.34(1)& 0.36(1) & 0.38(1)    \\  
\phantom{\Huge{l}} \raisebox{.1cm}{\phantom{\Huge{j}}}
$g_{B^*B\pi} f_{B^*}/m_{B^*}$   & 2.45(5)  & 2.10(4) & 1.90(4) & 1.76(4) & 1.65(4)  \\  
\phantom{\Huge{l}} \raisebox{.1cm}{\phantom{\Huge{j}}}
$\beta$   & 1.12(6)  & 0.19(7) & 1.26(8) & 1.32(9) & 1.39(10)  \\  
\phantom{\Huge{l}} \raisebox{.1cm}{\phantom{\Huge{j}}}
$m_{B^*_{0\ \mathrm eff}}/m_{B^*}$ & 1.06(3) &  1.09(3)  & 1.12(4) & 1.15(4)& 1.18(4)  \\  
 & & & &  & \\   \hline
\phantom{\Huge{l}} \raisebox{.1cm}{\phantom{\Huge{j}}}
$f_B /f_\pi$ & 2.5 $\pm$ 1.0 &  1.8 $\pm$ 0.6  & 1.6 $\pm$ 0.4& 1.4 $\pm$ 0.3 & 1.3 $\pm$ 0.3 \\  
 & & & &  & \\   \hline
\end{tabular}
\end{center}
\caption{\it Results of fits for a fixed $\gamma$. The last row is obtained
using Eqs.~{\rm (\ref{2nd})} and {\rm (\ref{f0})} in which we took $x=0.983$.}
\end{table}
We notice that the data tend towards larger values of the mass of the scalar
effective pole
as compared to the expected mass of the lowest scalar meson~\footnote{The scalar
 ($0^+$) meson is expected in the region of
 the measured bump~\cite{PDG} of $B^{**}$ states: $m_{B^{**}}=5.73$~GeV.},
  $\beta \simeq 1.17$, though they are consistent with this value.  Note 
 that an expected mass of the first excited vector state according to
 theoretical estimates should be close to 6 GeV,
 which again corresponds to a smaller value of $\gamma \approx 1.3$ (or
 $\alpha\approx 0.7$), but within the errors agrees with the values found
 in our analysis.  

\begin{figure}
\vspace*{-1.2cm}
\centerline{\epsfxsize12.0cm\epsffile{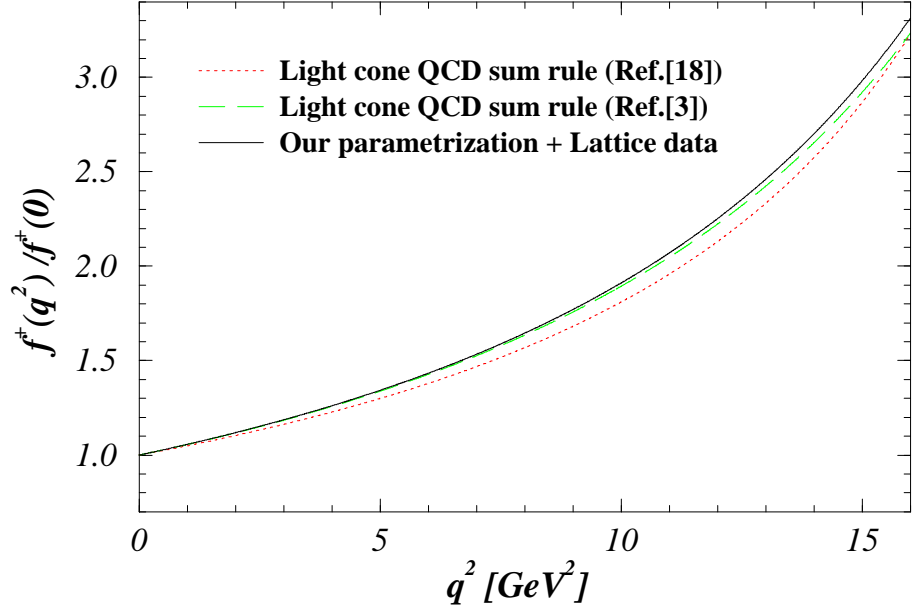}}
\vspace*{-0.1cm}
\caption[]{\it 
The shape of the form factor from our parametrization~{\rm (\ref{PARAMETR})}, 
constrained by the lattice data at
large $q^2$, is compared to the QCD sum rule predictions  at ``low'' $q^2$ {\rm (} Eq.{\rm (132)} in Ref.~{\rm \cite{QCDSR}}  for the dashed curve, and  Eq.{\rm (23)} in Ref.~{\rm \cite{Baganetal}} for the dotted curve{\rm )}. A very good agreement is found.
}
\label{FIG3}
\end{figure}

Now, when we have fixed the parameters by the lattice data~(\ref{RESULT}), we would like 
to compare the resulting shape of the form factor $f^+(q^2)$ to the result  of the 
light cone QCD sum rule, which
is applicable in the low $ q^2$ region.
To do that we define $f^+(q^2)/f^+(0)$, and
plot them both in Fig.~\ref{FIG3}. We see that our parametrization, 
constrained
by the lattice data at large $q^2$'s, indeed reproduces the shape
of the form factor at low $q^2$'s as predicted by QCD sum 
rule.
To be consistent, in the comparison we used the physical $m_{B^*}=5.32$ GeV. 

\vspace{1.cm}

\noindent
{\bf 4.~\underline{Summary}}~~
In this letter we proposed the physically more informative 
way to fit the data for the form factors in heavy to light 
decays. This is illustrated on the example of $B\to \pi$ 
semileptonic decay, and can be also used in  
$D\to \pi$, as well as in the other {\sl heavy $\to$ light} transitions. Initially, the parametrization for two 
form factors contains four parameters~(\ref{PROP}),(\ref{f0}), which we further 
reduced to 
three~(\ref{PARAMETR}). 
All parameters scale with the heavy mass as 
${\footnotesize {\mathrm const.}}\, + {\cal{O}}(1/m_B)$, except for $c_B$. 
The proposed 
parametrization gives us the information on the singularities 
which decisively influence the shape of the form factor in the 
semileptonic region, and allows us to extract the value of the 
residue at the nearest pole ({\it i.e.} to determine 
the value of $g_{B^*B\pi}$-coupling). 

On the specific example, we have shown that when fixing the
parameters in our parametrization by the lattice data in the
large $q^2$ region, one reproduces (to a good precision) the shape
of the form factor obtained by QCD sum rules at low and intermediate
$q^2$'s. 

We believe that this parametrization 
may be useful in the forthcoming analyses of systematically 
improved lattice data, as well as in experimental studies of exclusive
{\sl heavy $\to$ light} decays.

\vspace{1.7cm}
\noindent
{\bf Acknowledgement.}~~ We are very grateful to J.~Charles, G.~Martinelli and
O.~P\`ene for valuable comments on the manuscript, and to H.~Matufuru for correspondence. We also thank P.~Ball for important remarks.

\vspace{1.5cm}

\newpage

\end{document}